# Understanding Fission Gas Bubble Distribution, Lanthanide Transportation, and Thermal Conductivity Degradation in Neutron-irradiated α-U Using Machine Learning

Lu Cai[1*]; Fei Xu[3*];

Fidelma Dilemma[1]; Daniel J. Murray[1]; Cynthia A. Adkins[1]; Larry K Aagesen Jr[1].

Min Xian[2,#]; Luca Caprriot[1#]; Tiankai Yao[1#]

[*]Those authors contribute equally

[#]Corresponding authors

Min Xian: Email: mxian@uidaho.edu

Luca Caprriot: Email: Luca.Capriotti@inl.gov

Tiankai Yao, Email: tiankai.yao@inl.gov, Cell: 518-308-9256; Work: 208-533-7365

**Abstract:**

UZr based metallic nuclear fuel is the leading candidate for next-generation sodium-cooled fast reactors in the United States. US research reactors have been using and testing this fuel type since the 1960s and accumulated considerable experience and knowledge about the fuel performance. However, most of knowledge remains empirical. The lack of mechanistic understanding of fuel performance is preventing the qualification of UZr fuel for commercial use. This paper proposes a data-driven approach, coupled with advanced post irradiation examination, powered by machine learning algorithms, to facilitate the development of such understandings by providing unpreceded quantified new insights into fission gas bubbles. Specifically, based on the advanced post irradiation examination data collected on a neutron irradiated U-10Zr annular fuel, we developed a method to automatically detect, classify ~19,000 fission gas bubbles into different categories, and quantitatively link the data to lanthanide transpiration along the radial temperature gradient. The approach is versatile and can be modified to study different coupled irradiation effects, such as secondary phase redistribution and degradation of thermal conductivity, in irradiated nuclear fuel.

1. Introduction

U-10wt.% Zr (UZr) based metallic fuel is the leading candidate for next generation sodium cooled fast reactor due to low fabrication cost and capability to go very high burnup, over other fuel candidates. Although UZr fuel successfully powered United State research reactors, such as Experimental Breeder Reactor II (EBR II) and Fast Flux Test Facility (FFTF) [1], it has not been qualified for commercial use yet. Currently, Idaho national laboratory (INL) is leading the effort to support the qualification of UZr based fuel for commercial use. One route is to re-investigate the irradiated fuel pins from EBR II and FFTF through advanced characterization techniques, such as focused ion beam (FIB) sampling, transmission electron microscopy (TEM) characterization [2] and local thermal conductivity microscopy (TCM) measurement. The aim is to develop a mechanistic understanding of the evolution of and cladding microstructural, phase, and thermomechanical property evolution during irradiation. The ultimate goal is to provide a connection between the documented fuel performance data and fuel performance modeling tools, such as BISON, to facilitate fuel qualification at US Nuclear Regulatory Commission.

With the designed low smear density and ample plenum on fuel pin top, the burnup limitation for current UZr metallic fuel design is largely determined by the extend of fuel cladding chemical interaction (FCCI). The formation of eutectic compounds between lanthanide, a fission product, and Fe in cladding leads to the formation of eutectic compounds with low melting point that leads to significantly wastage of cladding materials that threatens fuel integrity and safety. How the lanthanide moves through fuel and arrives at fuel cladding interface is therefore critically important for a mechanistic understanding of UZr fuel performance. Existing theory supports the transportation of lanthanide through the interconnected bubbles from hot fuel center to cold cladding along the temperature gradient. The evolution of thermal conductivity of fuel during irradiation, therefore, is another important factor that dictates the movement of lanthanide and ultimately FCCI.

Unlike oxide nuclear fuel whose thermal conductivity degrades with irradiation induced defects and temperature, thermal conductivity of metallic fuel is immune to radiation induced defects. The radiation induced defects primary interference with phonons but have negligible effect on electros on which the thermal energy transport of metal materials relies on. For example, the thermal conductivity of Ni steel, used as light water reactor pressure vessel, even increases slightly after irradiation to 2.4E1023 neutron/m2. On the other hand, an in-pile measurement of thermal conductivity of metallic fuel indicates a 35% reduction, possibly due to the burnup induced gaseous swelling and sodium logging (Theodore H. Bauer 1995 nuclear technology 110 (1995)3). Compared with radiation induced defects, the fission gas bubbles have a much bigger impact on the thermal conductivity of metallic fuel.

Fission gas bubbles show different size and distribution pattern along radial thermal gradient inside irradiated UZr fuel. bubbles in cold rim zone close to cladding (Di Yun 2014 JNM paper) is large and of highly irregular shape, while round and relatively small in hot center zone. The classical approach to study fission gas bubble largely relies on imaging processing based on contrasts or manual select and measure of bubbles of interest. Those technique often carry human bias during measurement. Furthermore, advanced characterization method is producing large amount of data with bubble number beyond man's power to handle. Therefore, there is a strong need to develop a method that can accurately predict the fission gas bubble size, number density, and connectivity in automated manner, and develop physical descriptors to capture their effect on lanthanide movement and fuel thermal conductivity without human bias.

In this paper, we provide a framework with four steps to extract statistic data of fission gas bubbles from a scanning electron microscopy images size to 7,000 × 7,000 pixels. Traditional image processing methods were used to segment bubbles automatically in the first two steps. In the third step, we classifiy all bubble into three types: connected bubbles with lanthanide nodules; connected but empty bubble; and isolated bubbles. Based on the observation of features for the three types of bubbles, we developed a machine learning model to distinguish the bubble types. The model employed 18 features of bubbles, such as bubble's mean intensity, size, intensity standard deviation, intensity histogram, intensity range and the shape convexity. A data driven machine learning approach, named as Decision Tree, is employed to generate a bubble classifier and validated against a manually annotated dataset with 769 bubbles. Once trained, bubble detection and classification were applied across the entire fuel cross section to derive bubble statistic along radial temperature gradient.

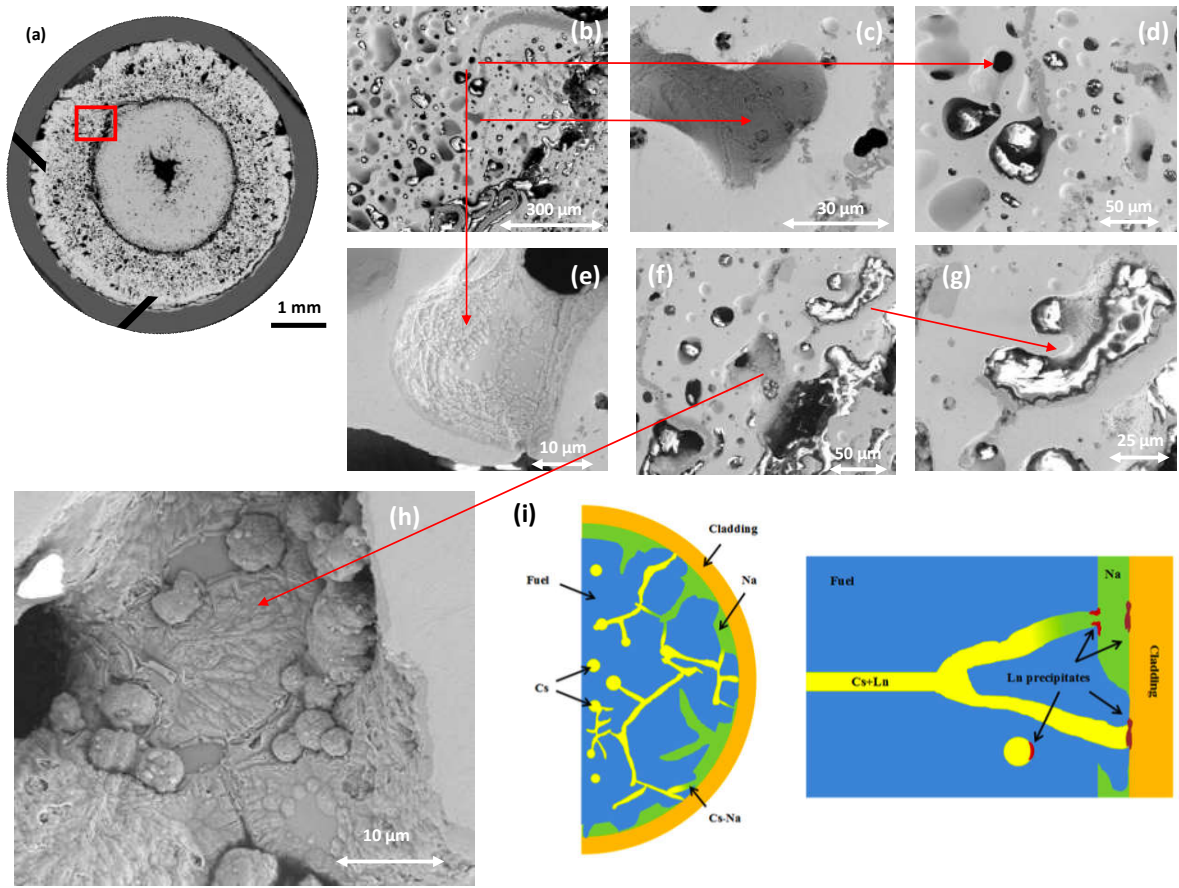

Fig. 1. The cross section of an U-10Zr irradiated to 3.4% FIMA burnup (a) and different bubble types (b-h) in the outer ring zone of α-U. One bubble type is important for lanthanide transportation, showing residence of lanthanide nodules (h) inside. A scheme of liquid like lanthanide moment (i) showing the short path provided by connected bubbles.

## 2. The proposed method

### 2.1. Experimental data

Readers are advised to read our previous publication for detailed advanced PIE study of this fuel cross section. In Fig. 1, a galley of bubbles inside α-U shows three different bubble types in Fig. 1(b),

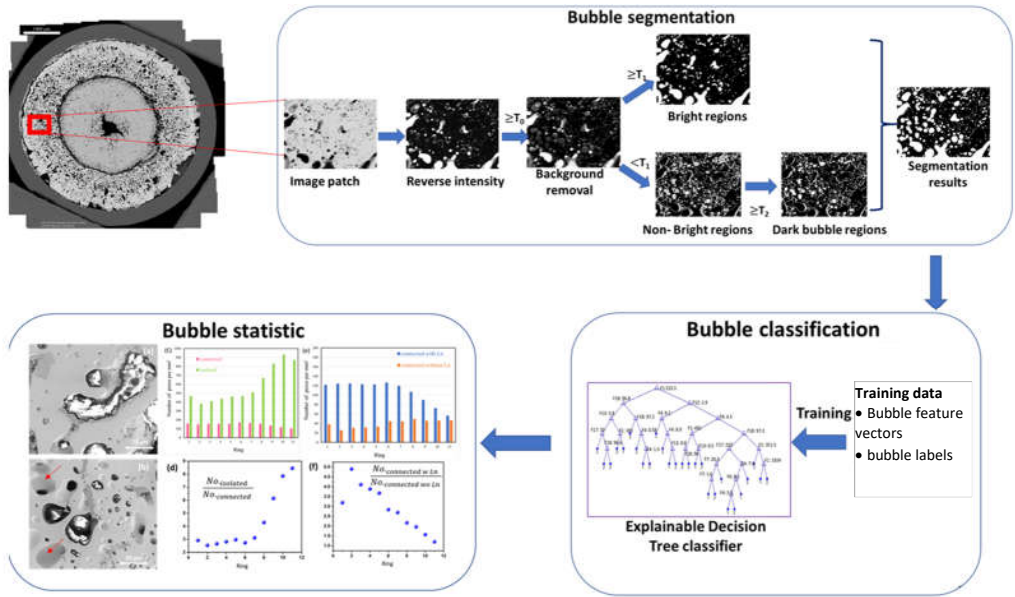

Fig. 2. Pipeline of the proposed approach.

Some bubbles as shown in Figs. 1(c), (e), and (h) show the attachment of nodules on bubble inner surface. Figure 1i provides a scheme of lanthanide (colored in red) transportation through a liquid-like movement through connected bubbles. Driven by temperature gradient, the lanthanides arrived at the fuel (colored in blue) cladding (colored in orange) interface with the help of liquid Na, the thermal bond, and Cs, a low melting fission product.

The pipeline of the proposed method is shown in Fig.1.

### 2.2. Background removal

In SEM images, we define the non-bubble pixels as *background* and the bubble pixels as *object*. As shown in Fig. 3(a), typically, the background appears brighter than the objects in the SEM images; and the bubbles have various shapes and sizes. We classify the bubbles into three categories: connected bubbles with fission product (white), connected empty bubbles (blue), and isolated bubbles (green), as shown in Fig. 3(b).

To make the bubbles more visible, firstly we reverse the image intensities by using the maximum intensity minus the current intensities. In the reversed image, the background appears darker than bubbles.

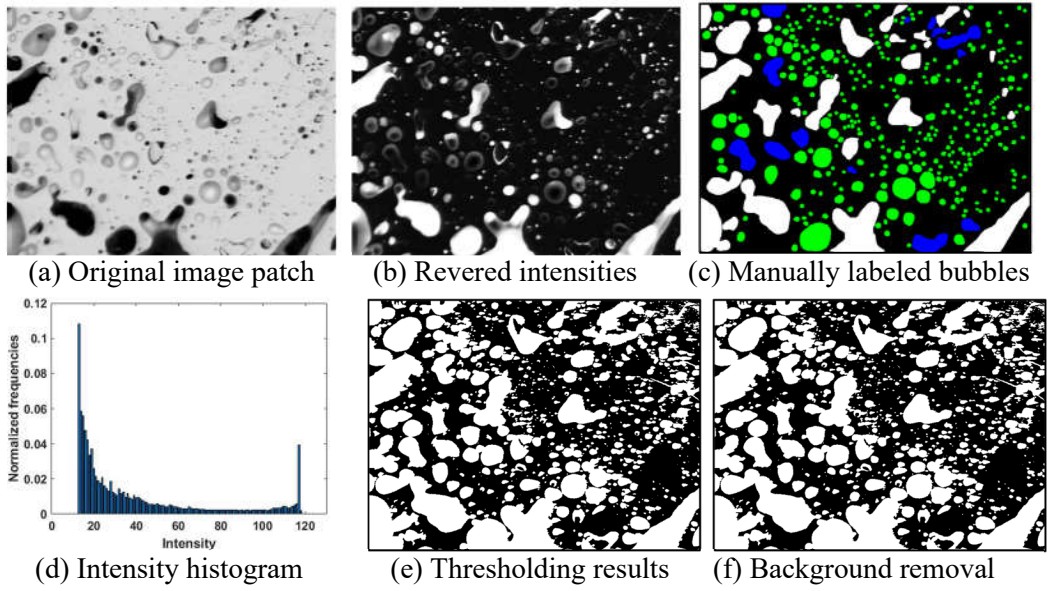

(a) Original image patch (b) Revered intensities (c) Manually labeled bubbles

(d) Intensity histogram (e) Thresholding results (f) Background removal

Fig. 3. A sample image patch and background removal.

Then the thresholding method is applied to remove background regions. Thresholding will generate a binary image (Fig. 3(c)), i.e., if the image intensities are less than a threshold $T_0$, the corresponding pixels assign 0s (background), otherwise assign 1s (bubble). One of the major challenge is to choose an appropriate threshold $T_0$ to separate background and bubbles. A large threshold may lose some bubbles partially, while too small threshold will result in high false positives. The threshold $T_0$ is chose by using the normalized histogram of pixels intensity values which is defined by

$$hist(k) = \frac{1}{N} \sum_{(i,j)} \delta(\lfloor I(i,j)/d \rfloor, k), k = 0, 1, \cdots, B-1 \tag{1}$$

$$\delta(\lfloor I(i,j)/d \rfloor, k) = \begin{cases} 1, & if\ k = \left\lfloor \frac{I(i,j)}{d} \right\rfloor \\ 0, & otherwise \end{cases} \tag{2}$$

where $N$ is the number of image pixels; $I(i, j)$ is the intensity of image pixel at location $(i, j)$ whose value is in [0, 255]; $\lfloor I(i,j)/d \rfloor$ denotes the floor division; $d$ is the intensity interval which is set to 2; $\delta(\cdot)$ is an indicator function that accumulates pixel for each bin; and $B$ is the number of bins of the histogram and is set as $\lfloor 255/d \rfloor + 1$. Therefore, the histogram defines the normalized frequencies ($\in$[0,1]) for all intensity values in each bin and is a simple and effective approach to describe image intensity distribution.

As shown in Fig. 3(d), there is a major peak between bins 0 to 20 which corresponds to the majority background pixels. Therefore, to remove the background pixels, the optimal threshold should be larger than the intensity of the peak. We set $T_0$ to the intensity of the first peak. A peak is defined if its value is larger than the values of the three previous bins and the next three bins. Fig. 3(f) shows the background removal result of a sample image patch.

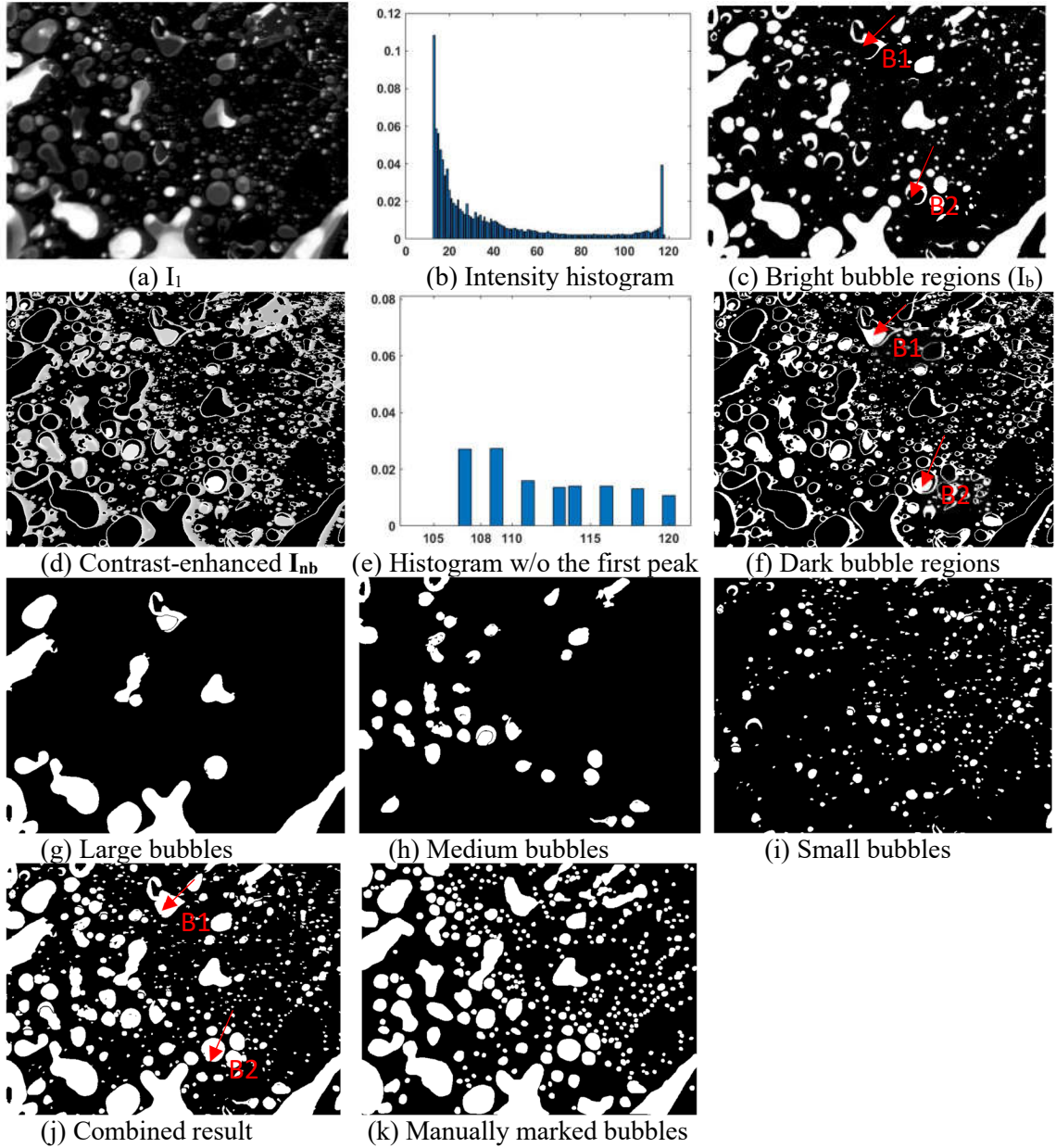

Fig. 4. Bubble segmentation using automatic thresholding.

### 2.3. Bubble segmentation using automatic image thresholding

Bubble segmentation aims to extract fission gas bubbles from SEM image. The background removal step excluded most background pixels, and this section is to locate all bubbles and their boundaries using the output image ($I_1$) of Section 1. In the original SEM images (Fig. 3(a)), the gas bubbles have various sizes and shapes, and have both bright and black regions. Therefore, it is difficult to extract all bubbles using one single intensity threshold; and we proposed a double-threshold strategy to segment both dark and bright bubble regions automatically. The first threshold ($T_1$) is applied to segment the bright regions, and a second threshold $T_2$ is applied to extract dark bubble regions. $T_1$ is chose as the first valley of the histogram of intensities of $I_1$. A valley point has smaller value than its two neighboring bins. $T_1$ separate pixels in $I_1$ into two parts, bright regions ($I_b$) and non-bright regions ($I_{nb}$). The contrast of $I_{nb}$ is enhanced by linearly mapping its pixel intensities to [0, 255] before thresholding. $T_2$ is chose as the first valley after the second peak in the intensity histogram of the contrast enhanced $I_{nb}$, because the first peak is at bin 0 and refers to the large number of background pixels; and dark bubble regions correspond to intensities after the second peak.

Based on the observation (marked red arrows in Figs. 4(c) and 4(f)), a bubble could have both bright and dark regions, and these regions might be generated at different steps. Therefore, we develop a bubble size-aware strategy to combine the regions into one bubble. First, all dark bubble regions are put into four groups, large (> 2000 pixels), medium (500 - 2000 pixels), small (50 – 200 pixels) and extra small (10 - 50 pixels). Then morphological opening operations are applied to the large, medium and small groups to disconnect neighboring bubble regions. Finally, the logic union operation will be applied to combine results from the above step and the $I_b$.

In Fig. 5, the proposed segmentation approach is applied to a new image patch. The first column shows a whole SEM image and its results; and the second column show an image patch (marked in the red rectangle) and the results. Images from rows one to five are the original image, reversed image, bright regions, dark regions, and final segmentation result, respectively.

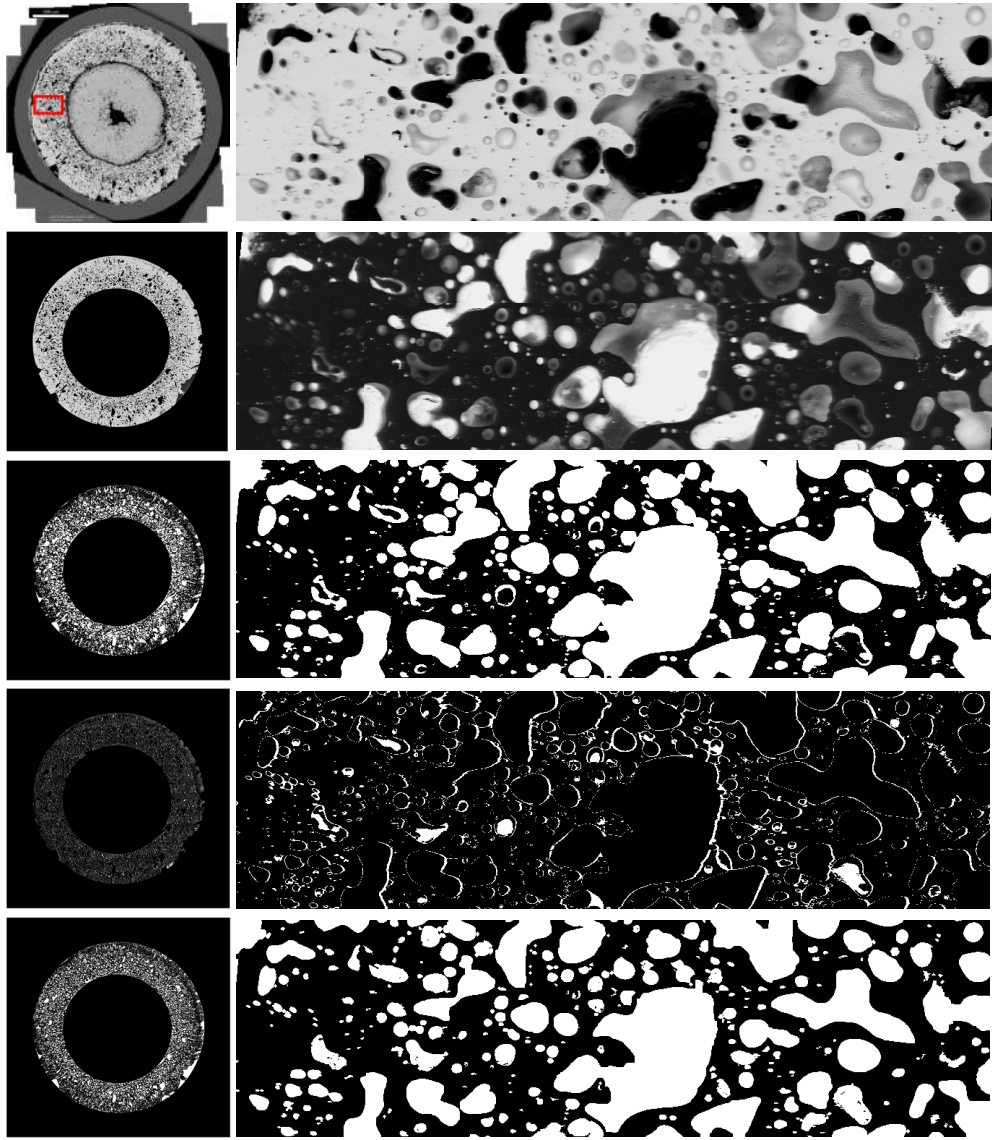

Fig. 5. Bubble segmentation for a whole SEM image.

### 2.4. Bubble classification using decision tree

For each bubble, we extract 18 features including the bubble size (**F1**), intensity histogram (**F2 - F14**), mean intensity (**F15**), intensity standard deviation (**F16**), intensity range (**F17**) and the shape convexity (**F18**). All bubbles are classified into three types, connected fission product bubbles ('**1**'), connected empty bubbles ('**2**'), and isolated bubbles ('**3**'). A decision tree (DT)-based machine learning algorithm is utilized to provide an interpretable bubble classification.

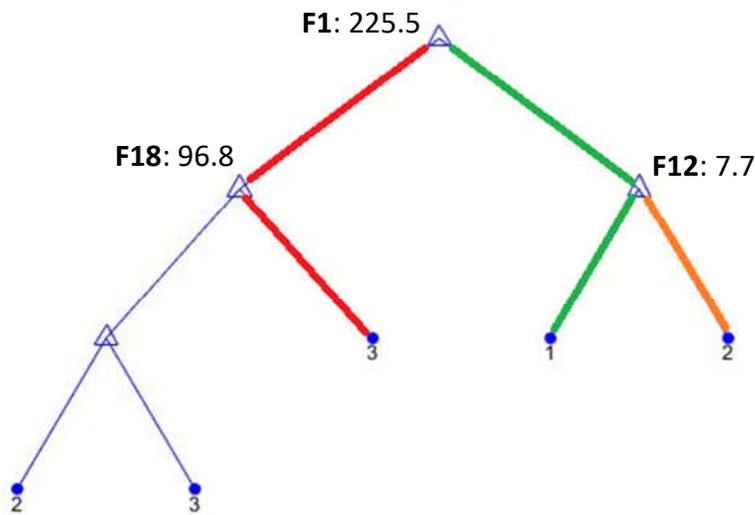

Fig. 6. A Decision Tree classifier.

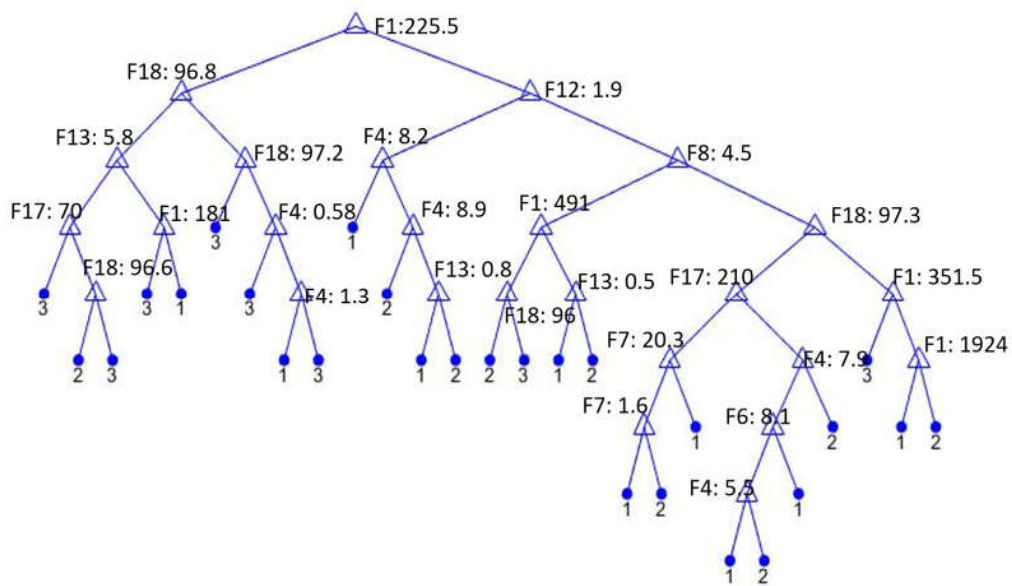

Fig. 7. The full decision tree for bubble classification.

The reasons for choosing DT are as bellow: 1) although artificial neural networks (ANNs), especially deep learning (DL), are widely used for classification problems, the approaches have poor interpretability. Meanwhile, DL-based methods need large, high-quality labeled training datasets to guarantee the performance, which might be unavailable in many applications. 2) The Linear Regression approaches are simple to implement and can also interpret the output efficiency. However, the predefined linear relationship among the features/properties may not suit many problems. Similarly. 3) DTs can model non-linear relationships and are highly interpretable. During the training process, a DT iteratively splits the feature space by choosing features and cut-points as the thresholds until the stopping criterion satisfied. The training/splitting process of a typical DT generates a binary tree, i.e., each tree node has two branches. For a given test sample, the decision-making process is to travel the tree from the root to one of the leaves. The process can be explicitly explained using the path the test sample visited. In addition, DTs can explain the importance of each input feature.

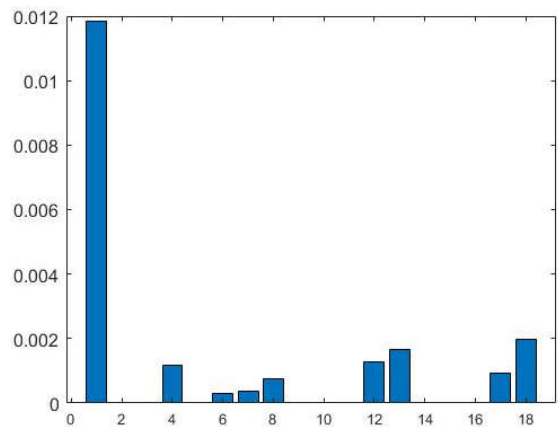

Fig. 8. Quantitative feature importance.

A sample binary DT is shown in Fig. 6. Each tree node is associated with a feaure and a cut-point to split the feature space, e.g., the root node applys cut-point value 225.5 to the feature **F1**. Each path from the root to a leaf is a decision-making process using a set of combined conditions. For an instance, the red path classifies a bubble into isolated bubble ('**3**') if the bubble's size (**F1**) is less than 225.5 and the convexity (**F18**) is greater than 96.8; if **F1** is greater than 225.5 and the 11$^{th}$ bin of the histogram (**F12**) is less than 7.7, the bubble will be classified into connected bubble with fission product ('**1**'); if **F1** is greater than 225.5 and **F12** is greater than 7.7, the bubble will be classified into connected empty bubble ('**2**'). The full traindecision tree is shown in Fig. 7. Meanwhile, the DT model outputs the importance of each feature. As shown in Fig. 8, **F1** is the dominat feature that determins the bubble types, and **F4, F12, F13, F18** are relatively important features.

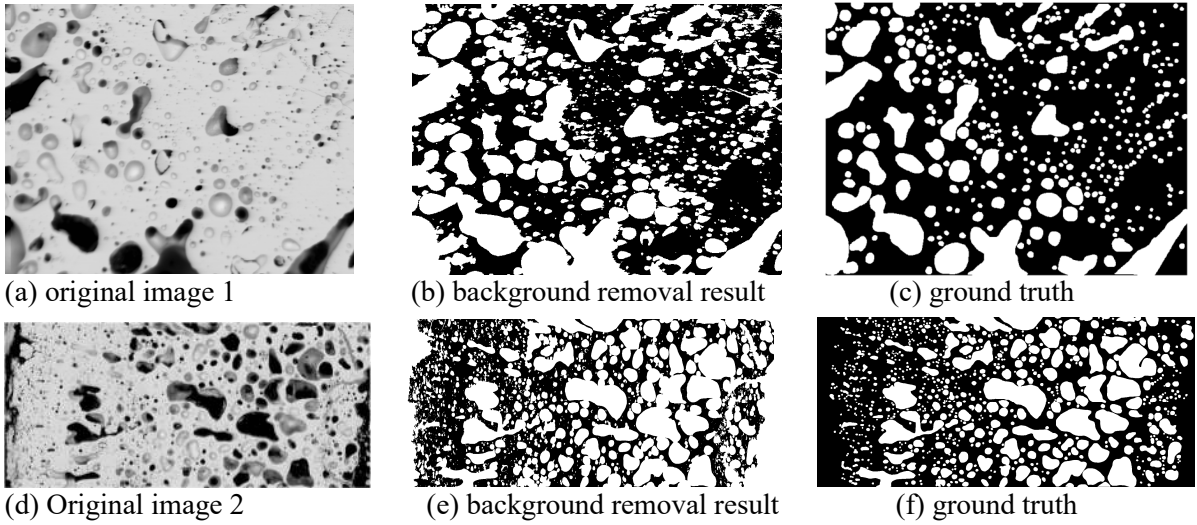

(a) original image 1    (b) background removal result    (c) ground truth

(d) Original image 2    (e) background removal result    (f) ground truth

Fig. 9. Background removal result.

## 3. Results and Discussion

### 3.1. Bubble segmentation evaluation

Evaluation of background removal. Since only partial annotated image is available, the ground truth (*GT*) are the pixels which are marked as bubbles. The following metrics are used to evaluate the performance: true positives ratio (*TPR*) and true negative ratio (*TNR*). *TPR* evaluates the ratio that the bubbles in the ground truth are detected as bubbles. *TNR* evaluates the ratio that the bubbles in the ground truth are not detected as bubbles. The metrics are defined as follows:

$$TPR = \frac{|GT \cap A_s|}{|GT|}$$

$$TNR = 1 - TPR$$

where $A_s$ is the segmentation map, and *GT* is the ground truth map. For all the metrics, the values are in [0,1], and the higher value of *TPR* and lower value of *TNR* indicate better performance. Segmentation results of two samples images are shown in Fig. 9. The *TPR* values of Fig.9(a) and Fig.9 (d) are **0.96** and **0.97**, respectively.

Evaluation based bubble detection and segmentation. In the paper, the commonly used metrics, Precision Ratio (PR) and Intersection Over Union (IOU), are utilized to evaluate the performance based on different bubble size level, respectively. IOU is a metric that finds the difference between ground truth annotations and predicted results. Precision ratio is to the number of correct detected objects divided by the number of ground truth annotations. For each bubble $i$ in the ground truth notations, we will match it by using the object $j$ in the detection result which has the maximum intersection set with the bubble $i$. The following criteria will be applied to determine whether the detection is correct or not: 1) bubble $i$ is a small size bubble and the intersection set contains at least 10 pixels; 2) bubble $i$ is a medium size bubble and the intersection set over the number of pixels of $i$th bubble is over 0.3; 3) bubble $i$ is a large size bubble and the intersection set over the number of pixels of ith bubble is over 0.5. The $IOU(k)$ is calculated by the following equation, where $k$ indicates the small, medium and large bubbles with value 1, 2, 3. The overall performance is listed as Table 1.

$$IOU(k) = \frac{\sum_{bubble\ i\ \in k} \frac{|bubble(i) \cap bubble(j)|}{|bubble(i) \cup bubble(j)|}}{\sum_{bubble\ i\ \in k} 1}$$

Table 1. Overall performance.

|  |  | Small bubbles (<=50 pixels) | Medium Bubbles (<=500 pixels) | Large Bubbles (> 500 pixels) |
|---|---|---|---|---|
| Fig.9 (a) | Number of bubbles | 8 | 295 | 53 |
|  | Detected bubbles | 6 | 258 | 48 |
|  | Precisions | 0.75 | 0.87 | 0.91 |
|  | IOU | 0.55 | 0.74 | 0.77 |
| Fig. 9 (d) | Number of bubbles | 19 | 578 | 156 |
|  | Detected bubbles | 16 | 461 | 132 |
|  | Precision | 0.8421 | 0.7976 | 0.8462 |
|  | IOU | 0.4709 | 0.5865 | 0.7362 |

From Fig.9 and Table 1, we found the proposed method obtained better performance on the images with lower bubble density distribution. The reason is that the high bubble density case will cause more unexpected connected bubbles which makes a poor performance on large size bubbles. Additionally, the overall performances on medium and large size bubbles are much better than that on small bubbles. The proposed method ignores the small bubbles with less than 10 pixels, and the morphological erosion operations will cause some small bubbles are eliminated. Moreover, the detection of bubble boundaries could be improved. In the future, we will annotate more images and make a benchmark ready for other supervised Machine Learning-based approaches.

### 3.2. Bubble classification evaluation

To classify the bubble types with a data-drive model, we marked 789 bubbles, in which there are 81 connected with fission product bubbles, 95 connected empty bubbles, and 613 isolated bubbles. We separated the entire dataset into a disjointly training dataset (80% of the whole dataset) and a test dataset (rest of the 20% data). During the training stage, we will apply a ten-fold cross-validation technique to create and evaluate the DT model. The advantage of ten-fold cross-validation is that all samples in the training dataset participate in the training stage, and each sample is used for validation exactly once. In other words, every iteration will have 90% and 10% training data as training and validation, respectively. The test stage is to provide an unbiased evaluation of the final model by using the test dataset. The commonly used metrics Accuracy, Precision, Recall and F_measure will be utilized to evaluate the performance.

$$Accuracy = \frac{\sum_j answer_j == predict_j}{\sum_j 1}$$

$$Precision = \frac{\sum_j answer_j == predict_j}{\sum_j predict_j == 1}$$

$$Recall = \frac{\sum_j answer_j == predict_j}{\sum_j answer_j == 1}$$

$$F\_measure = \frac{2 * Precision * Recall}{(Precision + Recall)}$$

where *answer* is the bubble label sequence, and *predict* is the predict result sequence generated by the DT model. For all the metrics, the values are in [0,1], and the higher value of $Accuracy$ and $F\_measure$ indicate better performance. The overall performance is shown in Table 2.

Since our training dataset is bias, which has small portion of connected fission product bubbles and connected empty bubbles, but large amount of isolated bubble, the classification performance on the isolated bubbles is much better than the other two's. To improve the classification performance on the connected bubbles, we will annotate more bubbles and build a benchmark with balanced samples for each type of bubble. In the future, we will utilize the DT model to label the bubble's category if the category is difficult to be determined manually.

Table 2. Classification evaluation.

|  | Bubble types | Accuracy | Precision | Recall | F_measure |
|---|---|---|---|---|---|
| Test dataset | Connected with fission product bubbles | 0.85 | 0.47 | 0.29 | 0.36 |
|  | Connected empty bubbles | 0.86 | 0.32 | 0.32 | 0.32 |
|  | Isolated bubbles | 0.89 | 0.96 | 0.90 | 0.93 |
| Entire dataset | Connected with fission product bubbles | 0.94 | 0.69 | 0.84 | 0.76 |
|  | Connected empty bubbles | 0.94 | 0.75 | 0.74 | 0.74 |
|  | Isolated bubbles | 0.95 | 0.98 | 0.96 | 0.97 |

### 3.3. Bubble statistics

The bubble statistics of a total of ~20,000 bubbles over the area of 26 mm2 are shown in Figs. 10-12. Fig. 10(a) is an overview of fuel cross section and the illustration of 11 ring diagrams in annular α-U region numbered in a sequence, from Ring 1 close to the interface of annular α-U region and UZr2 + x center region to Ring 11 next to cold cladding inner surface. Fig. 10(b) shows the bubble density (the number of bubbles per unit area) at each of the 11 rings, with three categories of bubble size. For large bubbles (area coverage > 500 µm2 or diameter > 50 µm, suppose the bubbles are round shape), the bubble density stays relatively constant from Ring 1 to Ring 8, then decreases almost linearly to Ring 11. For the intermediate bubbles (50 < area ≤ 500 µm2 or 5 < diameter ≤ 50 µm), the bubble density stays relative stable from Ring 1 to Ring 8 but increases linearly to Ring 10 and decreases slight at Ring 11. The small bubble density follows a similar trend as the intermediate bubble density.

Though visually there seems a greater number of bubbles at outer rings, the actual porosity area is much higher at inner rings (close to the fuel center) than that at outer rings as shown in Fig. 11(a). The ratio of the solid area over the bubble area in Ring 11 is about 2.7 times of that in Ring 2, or 3, or 4 (Fig.11(b)). The thermal conductivity of the bubbles should be equal to that of the filling gas (helium or fission gas or mixed of them), which is much smaller than that of α-U (the solid area), so the inhomogeneous degradation of thermal conductivity along the radial direction is revealed by the ratio of each ring zone covered by solid vs bubble in area percentage, with inner rings much lower thermal conductivity and outer rings higher thermal conductivity. The bubble statistics, which has never been reported before and cannot be arrived without ML models, can be fed into fuel performance code, and make it possible to quantify the inhomogeneous thermal conductivity degradation from fuel center to cladding.

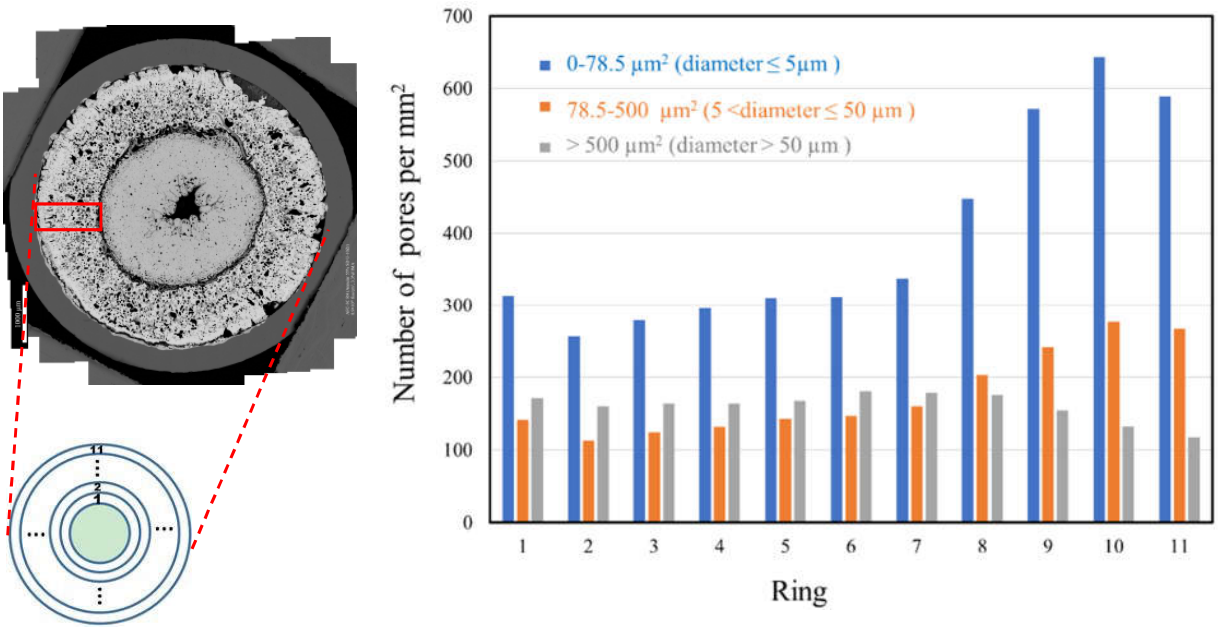

Fig. 10. (a) overview of fuel cross section and the 11 ring diagrams of α-U region numbered in a sequence. (b) the number of bubbles per unit area (bubble density) at each ring.

Lanthanides, fission products in the fuel, can react with the HT9 cladding and results in the thinning the cladding wall thickness and lowering the cladding melting point. Lanthanides moves down the temperature gradient along the interconnected bubbles (channel) to the inner cladding surface. As shown in Fig. 12(a), bubbles get interconnected due to irradiation and serve as a short path for lanthanides transportation. Not every interconnected bubble has lanthanides as shown in Fig. 12(b). It is necessary to train the ML model to distinguish isolated bubbles, connected bubbles, and connected bubbles with/without lanthanides for better understanding of lanthanide migration. As clearly demonstrated in Figure 12(c) and (d), the isolated bubble density (number per mm2) is much higher at outer rings (Ring 8-11) than at inner rings, proving that fission gas is easier to trap inside the fuel at lower temperature region.

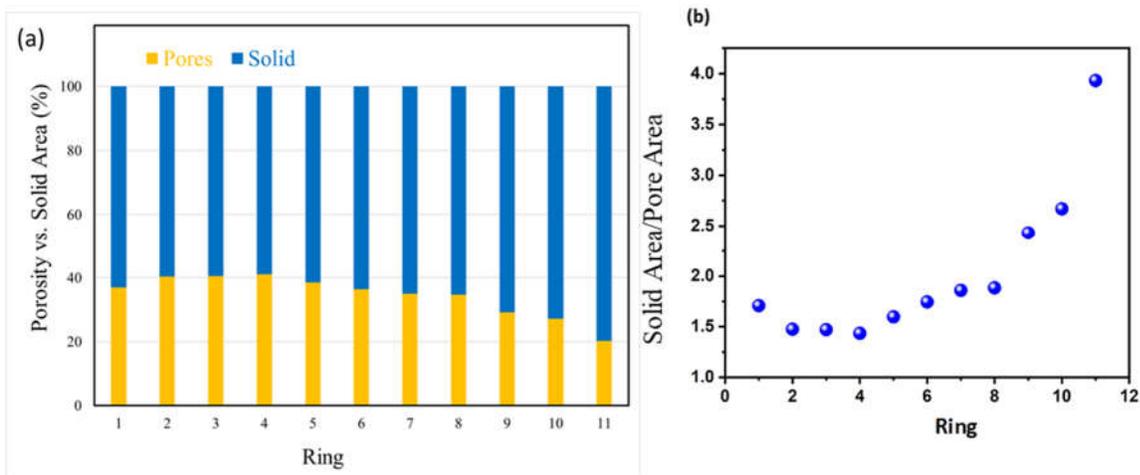

Fig. 11. (a) The percentage of porosity and solid area (b) the ratio of solid .at each ring.

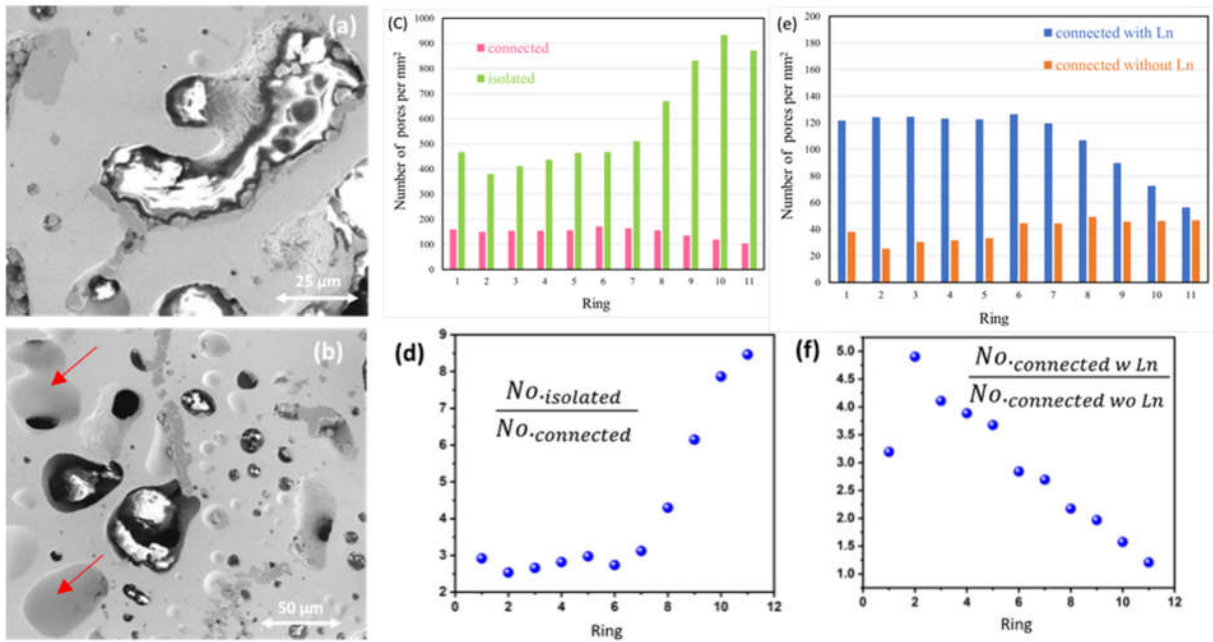

Fig.12. (a) image of connected bubbles with Ln inside, (b) image of connected bubbles do not have lanthanide at all. ((highlighted by red arrow), (c) the number of connected and isolated bubbles per mm$^2$ at each ring, (d) the ratio of number of isolated bubbles over that of connected bubbles at each ring, (e) within the connected bubbles, the bubble density with and without Ln at each ring (f) the ratio of number of bubbles with Ln over that without Ln.

Out of the connected bubbles, the bubble density with lanthanide stays relative stable from Ring 1 to Ring 7 and decreases almost linearly from Ring 7 to Ring 11 (Figure 12(e)). The connected bubbles close to the center is 5 times more likely to have lanthanide than outer fuel region (Figure 12(f)). How this bubble statistics relate to fuel performance needs further investigation, nevertheless, the ML provides us quantity data, which are extremely difficult to derive manually, to look into the lanthanide migration and FCCI.

4. **Conclusion**

This work presents a showcase of a machine learning approach to distinguish/identify the bubble features from microscope images of irradiated metallic uranium fuel. The annular α-U region with area of ~26 mm2 are split into 11 rings, within which about 20,000 bubbles are studied by machine learning. We are able to derive the bubble statistics and related it to fuel performance such as thermal conductivity degradation and lanthanide migration. This quantitative data from machine learning can feed into fuel design code for better prediction of fuel performance.

5. **Reference**

1. Carmack, W. J., D. L. Porter, Y. I. Chang, S. L. Hayes, M. K. Meyer, D. E. Burkes, C. B. Lee, T. Mizuno, F. Delage, and J. Somers. "Metallic fuels for advanced reactors." *Journal of Nuclear Materials* 392, no. 2 (2009): 139-150.
2. Yao, Tiankai, Luca Capriotti, Jason M. Harp, Xiang Liu, Yachun Wang, Fei Teng, Daniel J. Murray et al. "α-U and ω-UZr2 in neutron irradiated U-10Zr annular metallic fuel." *Journal of Nuclear Materials* 542 (2020): 152536.